\documentclass[aps,prl,twocolumn,showpacs,showkeys, superscriptaddress,floatfix]{revtex4}
\usepackage{graphicx}
\usepackage[utf8]{inputenc}
\usepackage{amsmath}
\usepackage{mathtools}
\usepackage{algorithm}
\usepackage[noend]{algpseudocode}
\usepackage{verbatim}
\usepackage{amssymb}
\usepackage{color}
\renewcommand\vec[1]{\boldsymbol{#1}}

\newcommand{\mean}[1]{\left\langle #1 \right\rangle}

\newcommand\Fig[1]{Fig.~\ref{fig:#1}}

\newcommand{\scrA}{\mathcal{A}}
\newcommand{\scrS}{\mathcal{S}}
\newcommand{\scrC}{\mathcal{C}}
\newcommand{\myto}{ \!\! \to \! }
\newcommand{\vecS}{\vec{S}}
\newcommand{\Break}{\State {\bf Break}}

\newcommand{\A}{\mathcal{A}}

\DeclarePairedDelimiter\floor{\lfloor}{\rfloor}

\begin{document}
\date{\today}\title{Clock Monte Carlo methods}
\author{Manon Michel}
\email{manon.michel@normalesup.org}
\affiliation{Centre de math\'ematiques appliqu\'ees, UMR 7641,\'Ecole Polytechnique, Palaiseau, France}
\affiliation{Orange Labs, 44 avenue de la R\'epublique, CS 50010, 92326 Ch\^atillon CEDEX, France}
\author{Xiaojun Tan}
\affiliation{Hefei National Laboratory for Physical Sciences at Microscale and Department of Modern Physics, University of Science and Technology of China, Hefei, Anhui 230026, China}
\author{Youjin Deng}
\email{yjdeng@ustc.edu.cn}
\affiliation{Hefei National Laboratory for Physical Sciences at Microscale and Department of Modern Physics, University of Science and Technology of China, Hefei, Anhui 230026, China}
\affiliation{CAS Center for Excellence and Synergetic Innovation Center in Quantum Information and Quantum Physics, University of Science and Technology of China, Hefei, Anhui 230026, China}

\begin{abstract}
  We propose the clock Monte Carlo technique for sampling each
  successive chain step in constant time. It is built on a recently
  proposed factorized transition filter and its core features include
  its O(1) computational complexity and its generality.  We elaborate
  how it leads to the clock factorized Metropolis (clock FMet) method,
  and discuss its application in other update schemes.  By grouping
  interaction terms into boxes of tunable sizes, we further formulate
  a variant of the clock FMet algorithm, with the limiting case of a
  single box reducing to the standard Metropolis method.  A
  theoretical analysis shows that an overall acceleration of
  ${\rm O}(N^\kappa)$ ($0 \! \leq \! \kappa \! \leq \!  1$) can be
  achieved compared to the Metropolis method, where $N$ is the system
  size and the $\kappa$ value depends on the nature of the energy
  extensivity.  As a systematic test, we simulate long-range O$(n)$
  spin models in a wide parameter regime: for $n \! = \!  1,2,3$, with
  disordered algebraically decaying or oscillatory
  Ruderman-Kittel-Kasuya-Yosida-type interactions and with and without
  external fields, and in spatial dimensions from $d \! = \! 1, 2, 3$
  to mean-field.  The O(1) computational complexity is demonstrated,
  and the expected acceleration is confirmed.  Its flexibility and its
  independence from the interaction range guarantee that the clock method
  would find decisive applications in systems with many interaction
  terms.
\end{abstract}


\maketitle

Markov-chain Monte Carlo methods (MCMC) are powerful tools in
many branches of science and engineering
\cite{Ceperley_1995,Frenkel_1996,Landau_2000,
  Opplestrup_2006,Rogers_2006,Robert_2001,Liu_2001,Glasserman_2004}.
For instance, MCMC plays a crucial role in the recent success of
AlphaGo \cite{AlphaGo}, and appears as a keystone of the potential
next deep learning revolution \cite{Neal_1996,Ghahramani_2015}. To
estimate high-dimensional integrals, MCMC generates a chain of random
configurations, called samples. The stationary distribution is
typically a Boltzmann distribution and the successive moves depend on
the induced energy changes.  Despite a now long history, the most
successful and influential MCMC algorithm remains the founding
Metropolis algorithm \cite{Metropolis_1953} for its generality and
ease of use, ranked as one of the top 10 algorithms in the 20th century
\cite{Dongarra_2000}.

The Metropolis algorithm has, however, two major limitations.  First,
nearby samples can be highly correlated and, around the phase
transition, the simulation efficiency drops quickly as the system size
$N$ increases.  Second, an attempted move requires calculating the
induced total energy change, leading to expensive computational
complexities of up to O$(N)$ for systems with long-range interactions.
This issue is also very acute in machine learning, where likelihood
evaluations \cite{Bardenet_2015} scale with the number of data points.

\begin{figure*}
\includegraphics[width=1.0\textwidth]{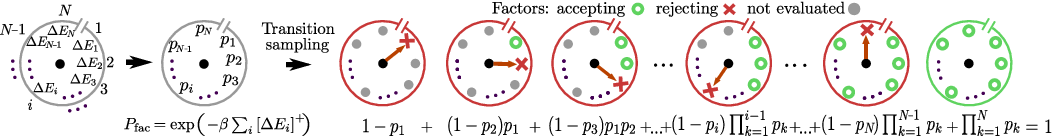}
\caption{Sketch of the clock sampling technique. The inhomogeneous
  Bernoulli process described by Eq.~(\ref{eq:Fact}) can be decomposed
  into $N\!+\!1$ \emph{clocks}, where the $i$th \emph{clock}
  ($i \! \leq \! N$) \emph{alarms} of the first occurrence of a
  rejection event at time $i$ and the $(N\!+\!1)$th \emph{clock}, with
  no alarm, represents an acceptance event. Such a process can be
  sampled within O(1) computational complexity.}
\label{fig:Clock}
\end{figure*}

Enormous efforts have been devoted to circumventing the two
limitations. Various efficient update schemes have been designed,
including the celebrated cluster and worm algorithms
\cite{Swendsen_1987,Wolff_1989,Prokofev_2001}, and the event-chain
(EC) irreversible method \cite{Bernard_2009,Michel_2014}.  Several
techniques are also available in reducing the computational complexity
for specific algorithms and systems.  An ``early-rejection"
  scheme was mentioned in the textbook~\cite{Frenkel_1996}, which is
  nevertheless of O$(N)$ complexity.  Making use of the particular
  feature that each bond is treated independently in the
  cluster-update scheme~\cite{Edwards_1988}, Luijten and
  Bl\"ote~\cite{Luijten_1995} applied an efficient sampling procedure
  to place occupied bonds, instead of visiting each bond sequentially
  and throwing a random number to decide its status.  The
  Luijten-Bl\"ote cluster algorithm has O(1)
  complexity~\cite{Luijten_1995, Fukui_2008, Flores_2017}, and has
  been generalized to quantum systems~\cite{Deng_2002}.  Recently, an
  EC algorithm was proposed for long-range soft-sphere systems
  \cite{Kapfer_2016}.

In this Rapid Communication, we propose a general ``clock" MC
  method, which has a constant-time sampling and can be applied to
  various update schemes. The core ingredient is the factorized
  Metropolis filter proposed in Ref.~\cite{Michel_2014}.  In
  particular, an algorithm of O(1) computational complexity is
  formulated in the framework of the local and most general update
  scheme, which we call the clock factorized Metropolis (FMet)
  algorithm. By grouping the interaction terms into boxes of tunable
  sizes, we further obtain a variant of the clock FMet algorithm for
  efficiency optimization. The limiting case of a single box recovers
  the Metropolis method, directly illustrating the generality of the
  clock FMet algorithm. We also discuss how the clock technique acts
  as a common ground for existing exact complexity reduction methods
  and present in particular its implementation in the EC update
  scheme.

  While gaining an O$(N)$ speeding-up in computational effort, the
  clock FMet algorithm can suffer from a lower acceptance probability
  than the Metropolis method. The {\em overall} acceleration comes
  from the compromise of these two effects. We provide a systematic
  performance analysis by classifying the system into three types of
  {\it strict, marginal} and {\it sub-extensivities}, and show that an
  overall acceleration can be achieved up to O($N$) for the strict
  extensivity and O($N^\kappa$) (1$>\! \kappa \! \geq$0) for the other
  two. As other exact complexity reduction methods belong to the
  algorithmic clock class, this analysis also applies to these
  techniques.  Finally, we extensively simulate long-range O$(n)$-spin
  models in a wide parameter regime: for $n \! = \!  1,2,3$, with
  disordered algebraically decaying or oscillatory
  Ruderman-Kittel-Kasuya-Yosida (RKKY)-type interactions, with and
  without external fields and in spatial dimensions from
  $d \! = \!  1, 2, 3$ to mean-field. The O(1) computational
  complexity is demonstrated, and the expected acceleration is
  confirmed. These achievements are based on the complementary
  combination of the factorized Metropolis filter, O(1) sampling
  procedures and the grouping trick.

  {\bf Clock FMet algorithm.} Consider a system described by a
  collection of states $\scrS$ with Boltzmann weights
  $\pi(\scrS) \propto \exp(-\beta E(\scrS))$, with
  $\beta = 1/k_{\rm b} T$ the inverse temperature.  The energy
  $E(\scrS) = \sum_i E_i(\scrS)$ is the sum of all interaction terms
  that are pairwise or more generally in many-body groups.  At each
  step, the Metropolis algorithm attempts to update a state $\scrS$
  into another $\scrS'$ with acceptance probability
\begin{equation}
 P_{\rm Met} = \min\left(1, \frac{\pi(\scrS')}{\pi(\scrS)}\right) =
 \exp(-\beta\left[\Delta E_{\rm tot}\right]^+ ) 
  \label{eq:Metro} 
\end{equation}
with $[x]^+ = \max(0, x)$.  Evaluating the induced energy change
$\Delta E_{\rm tot}\! \equiv \! \sum_{i} \Delta E_i$ requires a costly
computation of all the involved interactions.  Therefore, we focus now
on the factorized Metropolis filter~\cite{Michel_2014}
\begin{equation}
P_{\rm fac} = \prod_{i} p_i(\scrS \to \scrS') =
\prod_{i} \exp (-\beta\left[\Delta E_i\right]^+ ) \; ,
\label{eq:Fact}
\end{equation}
which also satisfies the detailed-balance condition
$ \pi(\scrS)p(\scrS \myto \scrS') = \pi(\scrS')p(\scrS' \myto \scrS)$.
Hereinafter we omit the dependence on $\scrS \myto \scrS'$ except in
case it hinders the clarity. The factorized filter is a key component
of the recent EC methods, as it allows one to extract interesting system
symmetries. On a more general level, the factorization of transition rates
can also play an important physical role in dynamical studies
\cite{Hucht_2009}.  A crucial feature of Eq.~(\ref{eq:Fact}) is the
consensus rule: As the transition probability $P_{\rm fac}$ is a
product of independent factors $p_i$, an attempted move is accepted
{\em only if} all the factors give permission
(Fig.~\ref{fig:Clock}).  This leads to a lower acceptance probability
in Eq.~(\ref{eq:Fact}) than in Eq.~(\ref{eq:Metro}).  However, we show
here how it plays a key role in designing the clock technique that
dramatically reduces the computational complexity from O$(N)$ to O(1),
greatly improving the {\it overall} performance.

Without loss of generality, we illustrate the clock FMet 
method in the example of a long-range O$(n)$ model of $N \! + \! 1$
spins, with the Hamiltonian
\begin{equation}
  \mathcal{H} = -c(N) \sum_{i < j} J_{ij}\vec{S}_i \cdot
  \vec{S}_j \; , \hspace{5mm} (|\vec{S}|=1)
  \label{eq:Hamiltonian}
  \end{equation} 
  with $\vecS$ unit vectors in $\mathbb{R}^n$.  For
  $n \! = \!  1, 2, 3$, one has the Ising, XY and Heisenberg models,
  respectively.  The coupling strength $J_{ij}$ depends on distance
  $r_{ij}$, and can be ferromagnetic ($J_{ij} \! > \! 0$),
  anti-ferromagnetic ($J_{ij} \! < \! 0$), or disordered.  There
  are in total $N(N\!+\!1)/2$ interaction terms.  The normalization
  constant $c(N)$, scaling typically in $1/N^\alpha$
  ($1 \! \geq \! \alpha \! \geq \! 0$), 
  is to ensure the energy extensivity, which, e.g., is $1/N $ for a
  mean-field ferromagnet but $1/\sqrt{N}$ for the
  Sherrington-Kirkpatrick model~\cite{Sherrington_1975,Kirkpatrick_1978}.  An attempted move
  is to flip or rotate a randomly-chosen spin $\vecS_j$. This leads
  to an energy change
  $-c(N)\sum_{i\neq j}J_{ij}(\vecS'_j-\vecS_j)\cdot \vecS_i$, which
  requests an O$(N)$ computation in the Metropolis
  algorithm. 

  A straightforward implementation of Eq.~(\ref{eq:Fact}) is as
  follows.  One orders the factor terms from $i \! = \! 1$,
  sequentially samples the rejection of each factor $i$ with
  probability $1-p_i$, and stops at the first-rejecting factor $i_{\rm
    rej}$; if no rejection is sampled until factor $N$, the move is
  accepted. This is analogous to
  an inhomogeneous Bernoulli process of rate $p_i$, as illustrated in
  \Fig{Clock}, where the $i$th clock, with probability
  $P_{\text{rej}}(i) \! = \! (1 - p_i)\prod_{k =1 }^{i-1} p_k $,
  represents the event for the $i$th factor to be first-rejecting.
  Instead of sequentially sampling each factor, one can also evaluate
  cumulative probability $F_k \! = \!  \sum_{k'=1}^{k}
  P_{\text{rej}}(k')$ and directly obtain the $i_{\rm rej} \! = \! i$
  value by solving $F_{i-1} \! < \! \nu \! \leq \! F_{i}$ with a
  single random number $ \nu \! \in \!  (0,1]$.  Nevertheless, the
  individual probabilities $p_k$ depend {\it a priori} on a local
  configuration $(\vec{S}_i,\vec{S}_j)$, and each move still requires
  an average number $\scrC \!\sim\! {\rm O}(N)$ of $p_k$-evaluations.

  To avoid these costly evaluations, we introduce a bound Bernoulli
  process with a {\it configuration-independent} probability
  $\widehat{p_i}$, so that $1-\widehat{p_i} \geq 1-p_i (\scrS \! \myto
  \! \scrS')$, as done in Ref. \cite{Luijten_1995}.  An actual
  rejection at a factor $i$ corresponds to a bound rejection once
  resampled with relative probability
\begin{equation}
   p_{i,{\rm rel}} = (1 - p_{i})/(1 - \widehat{p_i}) \;.
\label{eq:ActualRej}
\end{equation}
At each factor $i$, three events are possible: (${\rm A}_1$) bound
acceptance with $p_i^{{\rm A}_1} \! = \! \hat{p}_i$, ($\rm A_2$) bound
rejection and resampling rejection with
$p_{i}^{{\rm A}_2} \! = \!  (1-\hat{p}_i) (1- p_{i,{\rm rel}})$, and
(R) bound rejection and resampling acceptance with
$p_{i}^{{\rm R}} \! = \!  (1-\hat{p}_i) p_{i,{\rm rel}}$.  Sampling
the $i$th clock, i.e. the first-rejection at factor $i$, is then
replaced by sampling a random path of events (${\rm A}_1$) or
(${\rm A}_2$) for $ k \! \leq \!  i \! -\! 1$ and a first event
(R) at $i$, as described by
\begin{equation}
 P_{\text{rej}}(i) =  p_{i}^{{\rm R}} \prod_{k=1}^{i-1} ( p_{k}^{{\rm A}_1} + p_{k}^{{\rm A}_2} )\; .
\label{eq:Prej}
\end{equation}
As the bound $\hat{p}_i$'s are configuration-independent, the
  bound cumulative probabilities $\widehat{F}_i$ can be analytically calculated or tabulated.
  Initializing $i_{\widehat{\rm rej}} \! = \! 0 $, the next bound rejection
  $i_{\widehat{\rm rej}}$ is updated to $i$ by solving,
\begin{equation}
  \widehat{F}_{i-1} < \nu(1-\widehat{F}_{i_{\widehat{\rm rej}}})\!+\!\widehat{F}_{i_{\widehat{\rm rej}}} \leq  \widehat{F}_{i}\;,
\label{eq:SampleRejection}
\end{equation}
and the resampling is then applied.
This is done within an O$(1)$ complexity.
If no actual rejection occurs, i.e. event (${\rm A}_2$), 
the procedure is repeated until an event (R) (actual rejection) is sampled or until
  $\nu(1-\widehat{F}_{i_{\widehat{\rm rej}}})\!+\!\widehat{F}_{i_{\widehat{\rm rej}}}>\widehat{F}_{N}$ (actual acceptance).
The overall complexity $\scrC$ identifies now
  with the average number of attempted bound rejections 
  $\!\sim\! {\rm O}(\ln P_B/\ln P_{\rm Fac}) \!\sim\! {\rm O}(1)$ 
  if the  bound consensus probability $P_B \! =\! \prod\widehat{p}_i $ scales with $N$ as $P_{\rm Fac}$.
 For a homogeneous case $\widehat{p}_i \! \equiv \! \widehat{p}$,
  Eq.~(\ref{eq:SampleRejection}) reduces to
  $i = i_{\widehat{\rm rej}}\!+\!\floor*{1\! +\! \ln (\nu) /\ln
    (\widehat{p})}$, which can be easily adapted to inhomogeneous
  bound probabilities by ordering the factors increasingly with
  $\widehat{p}_i$ and by replacing $\widehat{p}$ by
  $\widehat{p}_{i_{\widehat{\rm rej}}+1}$ \cite{Shanthikumar_1985}.
 Alternatively, one can directly generate the whole list of bound rejection events
by the Walker method \cite{Walker_1977,Marsaglia_2004} or its
Fukui-Todo extension \cite{Fukui_2008}, and then sequentially apply the resampling.

Algorithm~\ref{Algo:ClockMet} summarizes a clock FMet method for a long-range spin system.
For Hamiltonian~(\ref{eq:Hamiltonian}), $\widehat{p}_k $ can be taken
as a function of distance $r_{ij}$ as  $\widehat{p} (r_{ij}) \! = \! \exp(-2\beta c(N) |J_{ij}|) $.
\begin{algorithm}[H]
\begin{algorithmic}[l]
\State Draw a random spin $j$ and a random move $\vecS_j \myto \vecS_j'$ 
\State $i_{\widehat{\rm rej}} \gets  0$  \Comment{{\it Sample bound rejections starting from $i_{\widehat{\rm rej}}$}}
\While{True}
    \State $i_{\widehat{\rm rej}} \gets i $ \Comment{{\it Next bound rejection $i$ given by Eq.~(\ref{eq:SampleRejection})}}
    \If{$i_{\widehat{\rm rej}} \!   >   \! N$} 
    \State $\vecS_j \gets \vecS'_j$ \Comment{{\it Move accepted}}
    \Break
    \Else \Comment{{\it Decide whether it is an actual rejection}}
         \State $p_{i_{\widehat{\rm rej}}, {\rm rel}} \gets $ Eq.(\ref{eq:ActualRej})
         \If{$\text{ran}(0, 1) \leq p_{i_{\widehat{\rm rej}}, {\rm rel}}$}
         \Break \Comment{{\it Move rejected}} 
  
    \EndIf
    \EndIf
\EndWhile
\end{algorithmic}
\caption{Clock factorized Metropolis (Clock FMet)}
\label{Algo:ClockMet}
\end{algorithm}

The clock method can be applied to any transition
  probability expressed as a product of independent factors, as the
  one proposed in \cite{Hucht_2009} for instance. However, the factorized
  Metropolis filter, in addition to a maximal acceptance rate factorwise,
  presents the following advantage. As each factor in Eq.~(\ref{eq:Fact}) can contain an {\em
    arbitrary} number of interactions, we introduce a variant of clock
  FMet algorithm in which the interactions are grouped into ``{\em
    boxes}" b of {\em tunable} sizes $B_b$, as
\begin{equation}
   P^{\text{Box}}_{\rm fac}\!=\!\prod_{b} \exp
  \left(-\beta\left[\sum_{i=1}^{B_b}\Delta E_{b_i}\right]^+\right).
  \label{eq:Box}
\end{equation}
It leads to new optimization possibilities (e.g. how to group the
interaction terms). If all the interactions are in a single box, one
recovers the standard Metropolis method.

The clock technique has two important ingredients: the consensus rule and the resampling.
Both the ingredients are {\it general}: 
They do not depend either on any specific configurations, or on factor ordering, 
or on energy functions, or on update schemes.
For systems in a continuous volume, as soft spheres, 
one can introduce a grid \cite{Kapfer_2016} to which the clock technique is applied.

{\it Generalization to other update schemes.}  We illustrate the
generality of the clock method by discussing its application in the EC
method for the O($n$) spin model with $n \! \geq \!
2$.~\cite{Michel_2015,Nishikawa_2015}.  With an auxiliary {\em
  lifting} variable $j$ that specifies the moving spin, the EC method
proposes to rotate infinitesimally its angle as $\phi_j \myto \phi_j
\! + \! {\rm d}\phi$.  Such a move is rejected by {\em at most} one
spin $i$, owing to a continuous derivation of Eq.~(\ref{eq:Fact}).
This yields $p_i \to 1-\lambda_i {\rm d} \phi$ and
$P_{\text{rej}}(i)\to\lambda_i {\rm d} \phi$.  For each factor $i$,
the rejection event, with distance $\delta_i \phi$, is thus ruled by a
Poisson process (PP) of rate $\lambda_i$, continuous derivation of the
standard Bernoulli process.  The spin $j$ is then rotated by the
minimum distance $\delta \phi = \min (\delta_i \phi) $, and the
associated factor becomes the moving spin, i.e. $j \myto i_{\rm
  min}$. For long-range interactions, evaluating $\delta_i \phi$ for
all the factors becomes costly.

To derive the clock method, we introduce a bound Poisson process of total rate
$\bar{\lambda} = \sum \bar{\lambda}_i \; ( \bar{\lambda}_i \geq
\lambda_i )$, evaluate a random bound rotation
$\delta \widehat{\phi} \! = \! - \ln \nu/\bar{\lambda}$,
 sample the rejecting bound factor $i_{\widehat{\rm min}}=i$ with probability
$ \bar{\lambda}_i / \bar{\lambda}$, and resample it as an actual
  lift $j$ to $i$ according to  $ \lambda_i / \bar{\lambda}_i$
  (Eq.~(\ref{eq:ActualRej})).
This comes down to the thinning method \cite{Lewis_1979}, already
applied for soft-sphere systems~\cite{Kapfer_2016} and for logistic
regression in machine learning~\cite{Bouchard_2016}.

 We also note that the cluster methods \cite{Swendsen_1987,Wolff_1989}
 factorize each interaction term independently as in Eq.~(\ref{eq:Fact}).
 The resampling procedures in the extended cluster algorithms for
   long-range interactions and for quantum spin systems  \cite{Luijten_1995,Fukui_2008,Flores_2017, Deng_2002}  can be understood as specific cases of the clock method.

 \begin{figure}
  \includegraphics[width=1.0\columnwidth]{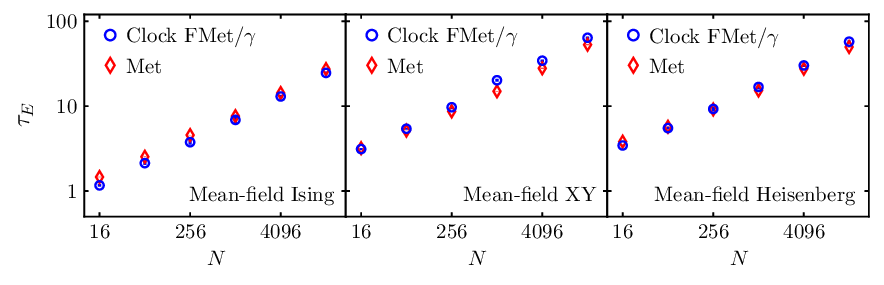}
  \caption{Integrated autocorrelation time $\tau_E$ of the energy for the
    ferromagnetic mean-field $O(n)$ model expressed in units of system
    sweeps. The values for the clock FMet algorithm are normalized by $\gamma$ (Eq.~(\ref{eq:RejFactor})).}
  \label{fig:Dynamics}
  \end{figure}

   {\bf Performance analysis.}  
   We expect and numerically confirm in Fig.~\ref{fig:Dynamics}
   that the standard Metropolis and the clock FMet algorithms have the
   same physical dynamics. The overall acceleration $\mathcal{A}$ in
   the latter comes then from the speeding-up in the complexity $\scrC$,
  corrected by the slowing-down $\gamma$ due to a lower
   acceptance in the factorized filter~(\ref{eq:Fact}), 
       leading to $\A \! \sim \! $ O$(N/\scrC\gamma)$.  Both effects can be
   characterized by the scaling of $\sum_i \max|\Delta E_i|$ and $\sum_i |\Delta E_i|$, as
   $\gamma = P_{\rm Met}/ P_{\rm Fac}$ can be written as
\begin{equation}
\ln \gamma  \! =\!
 \frac{\beta}{2} \left(\sum_i  |\Delta E_i |-|\sum_i \Delta E_i | \right) \; ,
  \label{eq:RejFactor}
  \end{equation} 
  and as $\scrC \! \sim \!  \ln P_B/\ln P_{\rm Fac}\!\sim\!\sum_i
  \max|\Delta E_i|/\sum_i|\Delta E_i|$.  Depending on the nature of
  the energy extensivity and phase of the system, the sum
  $\sum_i|\Delta E_i|$ may diverge as size $N \myto \infty$, while the
  sum $|\sum_i\Delta E_i|$ converges to a constant.  This normally
  occurs in disordered systems with slowly decaying interactions, in
  which the ``satisfied" and ``unsatisfied" interaction terms
  compensate each other. The divergence of $\gamma$ can be controlled
  by introducing enough compensation through boxes. For a
  constant size $B$, it increases the complexity to O($ B |\ln
  P_B|$), but leads to an acceleration $\sim\!{\rm O}(N/(B|\ln
  P_B|)$).  By definition, $B \! \propto \! N$ would ensure a maximal
  energy compensation but an O(1) acceleration.

    We classify the system into the three types of \emph{strict},
    \emph{marginal} and \emph{sub-} \emph{extensivities}, for which
    $\sum_i\max|\Delta E_i|$ respectively scales as O(1), O($\ln N$)
    and O($N^{\alpha}$) ($1 \! \geq \! \alpha \! > \! 0$).  We
    demonstrate that the clock FMet method of {\em tunable}
    constant box sizes $B$ might achieve an overall
    acceleration
\begin{itemize}
\item $ \mathcal{A} \! \sim \!$ O$(N)$ for strict extensivity,
  directly from $\gamma \! \sim \! $ O(1) and $\scrC\!\sim\! {\rm O}(1)$.
\item $\mathcal{A} \! \sim $O$(N^\kappa)$ (0$\leq  \!\kappa \! <$1) for sub-extensivity.  
  Depending on the phase, $\ln \gamma$ may diverge, up to $N^\alpha$. For
  the spin glass of algebraically decaying interaction as $1/r^\sigma$
  ($\sigma \!<\!1$), we find that a box size $B \propto N/N^{\omega}$,
  with a fine-tuning exponent $0 \! \leq \! \omega \! < \! 1$, gives a
  sufficient compensation and an O$(N^\kappa$)
 ($\kappa \!\sim\! [\omega-\alpha]^+$) acceleration.

\item $ {\rm O} (N/(\ln N)^2)\! \leq \! \mathcal{A}_{\rm margin}  \! \leq \! {\rm O} (N/\ln N) $ for marginal extensivity. 
   We observe that setting $B$ up to $\ln N$ can be necessary to control $\gamma$.
\end{itemize}

For frustrated systems, irrespective of which class they belong to,
 efficient cluster algorithms are normally unavailable due
to the huge cluster sizes.
 Given the substantial acceleration for all the three classes of {\em strict, marginal}, and {\em
  sub-extensivities}, the application of the clock FMet method is very promising.

{\bf Simulations.} 
We simulate three typical systems in statistical physics, including long-range Ising
spin glass, the disordered O$(n)$ model with random external fields, and the
O$(n)$ model with RKKY-type interactions.  We record the number of
energy evaluations $\scrC$ for each MC step, which for the
 Metropolis method is simply $\scrC \! = \! N$.  
We measure the integrated correlation times $\tau$ for magnetic 
susceptibility $\chi$ in the units of energy evaluations, and compute the
  overall acceleration $\scrA$ as the inverse ratio $\scrA \! = \! \tau_{\rm other}/\tau_{\rm FMet}$, 
  where $\tau_{\rm other}$ is for the Metropolis or the Luijten-Bl\"ote (LB) cluster method.
  For the Metropolis method, it comes down to $\scrA \! = \! N/(\gamma\scrC)$.

 \begin{figure*}
  \includegraphics[width=0.8\textwidth]{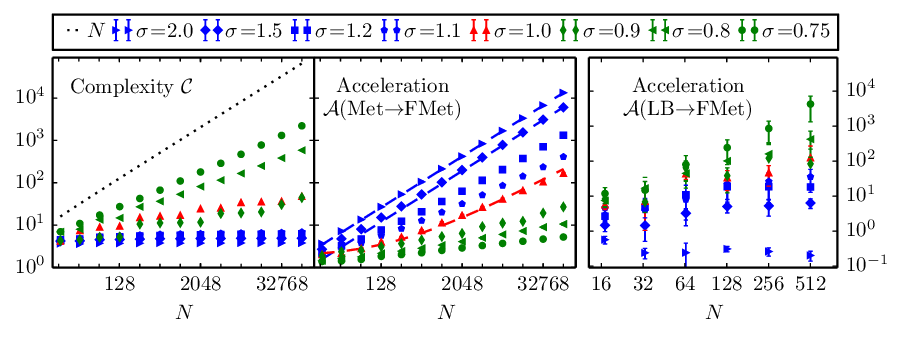}
  \caption{Complexity $\scrC$ (left) and acceleration
    $\mathcal{A}$ for the clock FMet algorithm for the 1D
    ($\beta \! = \! 1$) long-range Ising spin glass, compared to the Metropolis algorithm (middle) and the LB algorithm (right). The dashed blue and red lines respectively represent fits to $aN$ and $aN/(\log N)^2 + b$.}
  \label{fig:ComplexityLongRange}
  \end{figure*}

  {\it Long-range Ising spin glass.}  We consider a periodic
  one-dimensional (1D) spin glass defined by
  Eq.~(\ref{eq:Hamiltonian}).  The interactions decay algebraically as
  $J_{ij}=s_{ij}/r_{ij}^{\sigma}$ ($\sigma \! > \! 0$), with
  $s_{ij}=\pm 1$ from a bimodal distribution.  The normalization
  $c(N)$ is given by $c(N)^{-2} \!= \!  \sum_{j>1} \mean{J_{1j}^2}$.
  This system, with a tunable exponent $\sigma$, is particularly
  useful in revealing the crossover behavior from the low-dimensional
  to the mean-field spin glass~\cite{Kotliar_1983, Bray_1986,
    Beyer_2012}.  For simplicity, the simulation is made at the
  mean-field critical temperature $\beta=1$~\cite{Sherrington_1975}.
  Depending on the value of $\sigma$, we recover the three extensivity
  regimes, i.e. strict ($\sigma\!>\!1$), marginal ($\sigma\!=1$) and
  sub-extensivities ($\sigma\!<\!1$).  We group the interaction terms
  following respective $r_{ij}$ values and set the box size as $B=2
  \; (\sigma \! > \!  1)$, $\ln N \; (\sigma \! =\!1)$, and
  $N^{2(1-\sigma)} \; (\sigma \!<\! 1)$.  The results are shown in
  Fig.~\ref{fig:ComplexityLongRange}.  For $\sigma \! > \!1$, the
  computational complexity $\scrC$ converges to a constant, and a
  dramatic overall acceleration $\scrA \! \sim \! {\rm O} (N)$ is
  achieved.  For $\sigma \! = \! 1$, we have $\scrC \! \sim \! \ln
  N^2$ and a significant acceleration $\scrA$ converging to $N/(\ln
  N)^2$.  For $\sigma \! < \! 1$, both $\scrC$ and $\scrA$ increase
  sub-linearly as $N$.  The acceleration $\scrA$ drops as $\sigma$
  becomes smaller.  Nevertheless, given the simplicity of the clock
  FMet method, the gained improvement is still significant. We
    also compare its performances with the LB algorithm, which confirm
    the superiority of the local Clock FMet for disordered systems.
  These results are fully consistent with the performance analysis.

\begin{figure}
  \includegraphics[width=1.0\columnwidth]{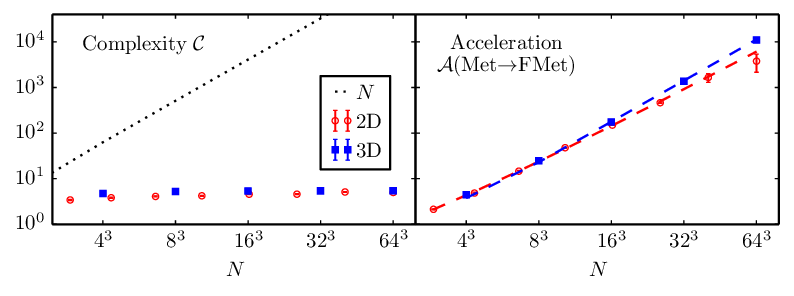}
  \caption{Complexity $\scrC$ (left) and acceleration $\mathcal{A}$ (right)
     for the 2D ($\beta \! = \! 1$, red circle) and 3D ($\beta \! = \!
    0.693$, blue square) RKKY Heisenberg spin systems, comparing the clock FMet
    method to the Metropolis algorithm. The dashed red and blue lines respectively represent fits to $aN$ and $aN/\log N+b$.}
  \label{fig:ComplexityRKKY}
  \end{figure}

  {\it RKKY-type interactions.}  We then consider the 2D and 3D
  Heisenberg models with oscillatory Ruderman-Kittel-Kasuya-Yosida
  (RKKY) interactions
  $J_{ij} \! = \! J_0 (\cos(2k_{\rm F} r_{ij})/r_{ij}^d)
  \exp(-r_{ij}/\lambda)$,
  where $d$ is the spatial dimension, $k_{\rm F} $ is the Fermi vector
  ($k_{\rm F} \! \approx \! 4.91$ for the spin-glass system $Cu$Mn), and
  $\lambda$ is the characteristic length in the damping
  term~\cite{Bray_1986,Matasubara_1992, Priour_2005, Priour_2006,
    Szalowski_2008,Kirkpatrick_2016}.  Due to their approximate
  description of real materials, rich behaviors, and important roles
  in bridging the experimental study of glassy materials and the
  spin-glass theory of short-range interactions
  \cite{Bray_1986,Matasubara_1992}, these systems are under extensive
  studies.  For simplicity, we set $ J_0 \! =\! 1$ and
  $ k_{\rm F} \! = \! \pi $, and take $\lambda \! = \! 3$ for 3D and
  $\lambda \! = \! \infty$ for 2D, so that the system is in the class
  of strict (3D) and marginal (2D) extensivities.  The simulations are
  at $\beta (2D) \! = \! 1$ and $\beta(3D) \! = \! 0.693 $, close to
  the critical temperature $\beta_c \! = \! 0.693 \, 003 (2) $ for the
  3D pure Heisenberg model~\cite{Deng_2005}.  Box sizes are set to 1
  and the achieved acceleration is again $\scrA \! \sim $ O$(N)$ for
  the strict extensivity, and $\scrA \! \sim $ O$(N/\ln N)$ for the
  marginal extensivity, as illustrated in Fig.~\ref{fig:ComplexityRKKY}.

\begin{figure*}
  \includegraphics[width=0.8\textwidth]{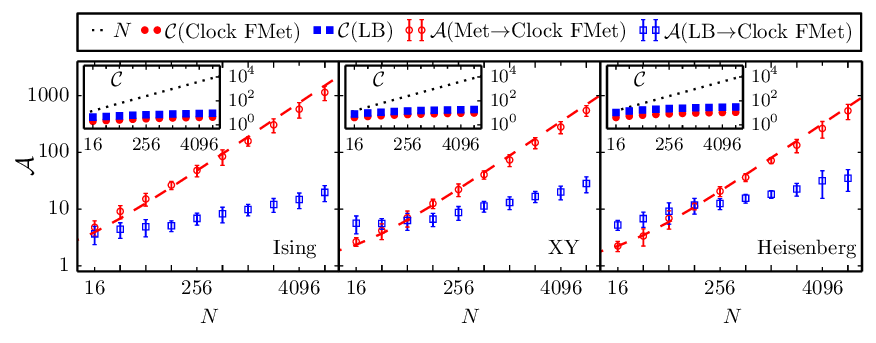}
  \caption{Acceleration $\scrA$ for the disordered mean-field O$(n)$
    model in random fields at $\beta \! = \! n$ (complexity
      $\scrC$ in insets).  The clock FMet method exhibits important
    acceleration $\scrA$, compared to both the Metropolis (red
    circles) and the LB cluster algorithm (blue squares). The dashed red line represents a fit to $aN+b$.}
  \label{fig:ComplexityE}
  \end{figure*}

  {\it Disordered random-field model.}  Finally, we study a disordered
  mean-field O$(n)$ model in a random external field.  The
  interactions are partly disordered, i.e. $J_{ij} \! = \! 1$ for 90\%
  of interactions while the remaining $J_{ij}$ are drawn from a normal
  distribution with $\langle J_{ij} \rangle \! = \! 0$ and
  $\langle J_{ij}^2 \rangle \! = \! 1$.  A quenched random field is
  applied to each lattice site as $-\vec{h_i} \! \cdot \!  \vec{S_i}$,
  where $\vec{h_i}$ is drawn from an $n-$dimensional normal
  distribution.  The normalization is $c(N) \! = \!1/N$, and the
  system belongs to the class of strict extensivity.  Random-field
  models have applications in a wide range of physics
  \cite{Larkin_1970,Efros_1975,Kirkpatrick_1994,Sethna_1993,
    Rosinberg_2008}, including the pinning of vortices in
  superconductors, Coulomb glass, the metal-insulator transition, and
  hysteresis and avalanche physics.  In general spatial dimensions,
  the thermodynamic properties and phase transitions are still debated
  \cite{Krzakala_2010, Leuzzi_2013}.  We perform simulations at the
  mean-field critical temperature $\beta=n$, with box sizes set to 1.
  The results are shown in \Fig{ComplexityE}.  The clock FMet method
  clearly displays an O$(N)$ acceleration over the Metropolis
  algorithm for all the Ising, XY and Heisenberg models.  It also
  exhibits some superiority ($\scrA \sim 50$ for large system sizes)
  compared to the LB cluster algorithm that already implements the
  clock technique and has an O(1) computational complexity.  The
  central-limit theorem tells that, as temperature is lowered and/or
  the strength of the external fields is increased, the acceptance rate
  exponentially drops for clusters of large sizes in the LB algorithm,
  and thus this superiority would become more pronounced.

  {\bf Conclusion.} We introduce a general clock technique with O(1)
  computational complexity for each Monte Carlo step, and discuss its
  implementations in various update schemes, regrouping most
    existing complexity reduction techniques into a single algorithmic
    class.  An important application is the clock FMet algorithm.
  This is made possible owing to the following three flexible
  features of the factorized filter~(\ref{eq:Fact}).
  First, the consensus rule in Eq.~(\ref{eq:Fact}) allows for the
  decision of the fate of a proposed move by an O(1) sampling
  procedure.  Second, the equal generality of Eqs.~(\ref{eq:Metro})
  and~(\ref{eq:Fact}) allows for a similar application range.  Third,
  the factorization range flexibility given by the grouping trick in
  Eq.~(\ref{eq:Box}) allows for a semi-continuation from the
  Metropolis Eq.~(\ref{eq:Metro}) to the factorized filter
  Eq.~(\ref{eq:Fact}) and a control over the frustration present in
  the considered system.

  The clock FMet algorithm and its variant with tunable box sizes can
  lead to significant or even dramatic acceleration $\scrA$.
  Depending on the system, theoretical analysis gives $\mathcal{A}$ up
  to O($N$), O($N/\ln N$) and O($N^\kappa$)
  ($1 \!  >\! \kappa \! \geq \!  0$) for respectively strict, marginal
  and sub-extensivities. Moreover, as the Metropolis method can be
  understood as a limiting case of the clock FMet, the latter cannot
  be worse. This is confirmed by simulations of long-range O$(n)$
  models in a wide parameter range. Since these systems are under
  active studies and the simulations rely heavily on the Metropolis
  method, the clock FMet algorithm is readily available to explore
  their rich physics. From its simplicity and ease
  of use, we conclude that the clock technique is a serious candidate
  for tackling Monte Carlo scaling in all scientific fields.

\section*{ Acknowledgments.}  XJT and YD thank the support by National
Natural Science Foundation of China under Grant No. 11625522 and the
Ministry of Science and Technology of China (under grants
2016YFA0301604), MM thanks the University of Science and Technology of
China for its hospitality during which this work was initiated and
partly done and is very grateful for the support from the Data Science
Initiative, the Chaire BayeScale "P. Laffitte" and the PHC program Xu
Guangqi (grant 41291UF).

\end{document}